\documentclass[aps,prl,
twocolumn,
nofootinbib,
notitlepage,
citeautoscript,
superscriptaddress,showkeys,showpacs]{revtex4-2} 
\usepackage{amsmath,amsfonts,amssymb}
\usepackage{color}
\usepackage[dvipsnames]{xcolor}
\definecolor{purple}{HTML}{900090}
\usepackage{tabularx}
\usepackage{graphicx}
\usepackage[colorlinks=true, allcolors = purple]{hyperref}
\usepackage[english]{babel}

\addto\captionsenglish{}

\newcommand{\sss}{\scriptscriptstyle}

\newcommand{\EF}{E_{\sss F}}
\newcommand{\kF}{k_{\sss F}}
\newcommand{\kB}{k_{\sss{B}}}
\newcommand{\muB}{\mu_{\sss{B}}}
\newcommand{\sinfty}{\sss \infty}

\newcommand{\inang}[1]{\langle #1 \rangle}

\newcommand{\m}[1]{
\bigg(
\begin{matrix}
#1
\end{matrix}
\bigg)}

\renewcommand{\vec}[1]{\boldsymbol{#1}}

\usepackage[figure,table]{hypcap}

\makeatother

\begin{document}
\title{A topological flux trap: Majorana bound states at screw dislocations}
\author{Stefan Rex}
\thanks{These two authors contributed equally}
\affiliation{Institute\,for\,Quantum\,Materials\,and\,Technologies,\,Karlsruhe\,Institute\,of\,Technology,\,Karlsruhe,\,Germany}
\affiliation{Institute for Theory of Condensed Matter, Karlsruhe Institute of Technology, Karlsruhe, Germany}
\author{Roland Willa}
\thanks{These two authors contributed equally}
\affiliation{Institute for Theory of Condensed Matter, Karlsruhe Institute of Technology, Karlsruhe, Germany}
\affiliation{Heidelberger Akademie der Wissenschaften, Heidelberg, Germany}

\begin{abstract}
The engineering of non-trivial topology in superconducting heterostructures is a very challenging task. Reducing the number of components in the system would facilitate the creation of the long-sought Majorana bound states.
Here, we explore a route toward emergent topology in a trivial superconductor without a need for other proximitized materials.
Specifically, we show that a vortex hosting an even number of flux quanta is capable of forming a quasi-one-dimensional topological sub-system that can be mapped to the Kitaev wire, if the vortex is trapped at a screw dislocation.
This crystallographic defect breaks inversion symmetry and thereby threads a local spin-orbit coupling through the superconductor.
The vortex-dislocation pair in the otherwise trivial bulk can harbor a pair of Majorana bound states located at the two surface terminations.
We explain the topological transition in terms of a band inversion in the Caroli-de~Gennes-Matricon vortex bound states and discuss favorable material parameters.
\end{abstract}

\maketitle

\section{1. Introduction}
Majorana bound states (MBS) are self-conjugate, non-Abelian quasi-particles that appear as boundary modes in topological superconductors \cite{Kit01, Iva01}. The foreseen applications of MBS in quantum computation \cite{NSS08} have motivated tremendous effort to generate them in a variety of systems, yet without conclusive evidence. To date, the design of Majorana platforms largely relies on the paradigm of proximity: Owed to the sparsity of intrinsic topological superconductors, non-trivial topology is engineered in heterostructures, e.g., in semiconductor nanowires \cite{ORO10, LSD10} with partial \cite{Mourik2012, Deng2012, Das2012, Rokhinson2012, Deng2016, Gul2018} or full \cite{Vaitiekenas2020} superconducting shell or at superconductors interfacing with topological insulators \cite{FuK08, XWL15}. The recurring key components are spin-orbit coupling, Zeeman splitting, and (trivial) $s$-wave superconductivity.

Here we take a different approach, where these components readily coexist without proximity effects and may cause the formation of a topological phase. The surprisingly simple system consists of a trivial superconductor in an external magnetic field, where a vortex is trapped at a screw dislocation. Screw dislocations are well understood and commonly occuring crystallographic defects, which introduce a spin-orbit coupling by locally breaking inversion symmetry \cite{Hu2018}.

While MBS at vortices usually require a $p$-wave superconductor \cite{Iva01}, we show below that this symmetry restriction can be lifted if the vortex carries multiple flux quanta \cite{Vaitiekenas2020, RGM19}. Depending on their type, superconductors tend to carry magnetic field either through large normal-state regions (type-I) or vortices with singly-quantized flux $\Phi_{0} \!=\! hc/2e$ (type-II). Yet, the formation of \emph{giant vortices}, carrying a flux $n \Phi_0$, $|n| \gtrsim 1$, is known to occur under appropriate circumstances \cite{Huebener2001}.
Generally, Caroli-de~Gennes-Matricon (CdGM) states \cite{Caroli1964} at sub-gap energies are found close to the vortex line. Including the effect of Zeeman splitting, an inversion of the CdGM bands is possible. We show that this inversion turns the vortex-dislocation system into an effective topological wire with MBS at the surface terminations. We thus demonstrate that, in principle, MBS can arise from a trivial superconductor, i.e., a superconductor where the 3D bulk does not possess nonzero topological indices in the absence of other adjacent materials.

We first introduce a minimal model and argue that MBS are symmetry-allowed for an even number of flux quanta. We then demonstrate the conceptual reduction of the system to a quasi-one-dimensional Majorana wire. Finally, we consider in more detail the spatial profile of multi-flux (giant) superconducting vortices and their CdGM states for vorticities $n>1$ and discuss suitable conditions to enter the topological regime.

\begin{figure}[tb]
\includegraphics[width = 0.45\textwidth]{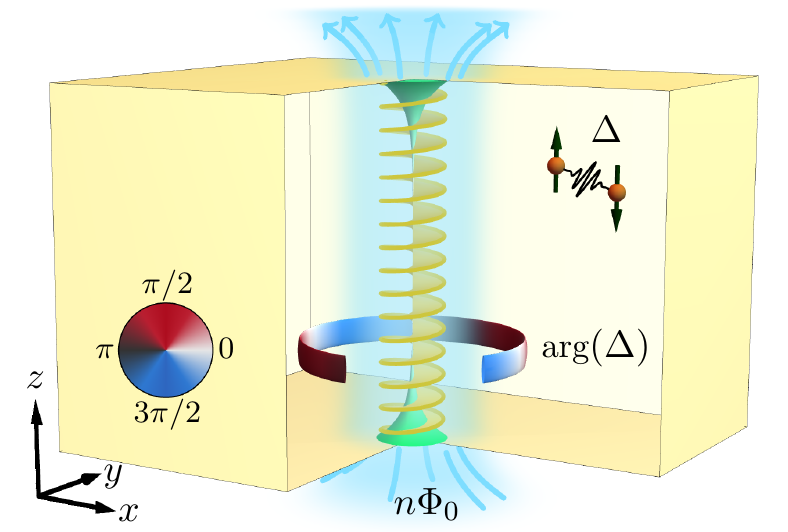}
\caption{
A trivial superconductor (yellow) is ready to host localized Majorana bound states (green) when a giant vortex (light blue) with even vorticity $n \in 2 \mathbb{Z}$ is trapped at a screw dislocation. The superconducting phase winds by $2\pi n$, indicated by the red/blue color scale.
\label{fig:schematic}
}
\end{figure}

\section{2. Model and symmetries}
Consider a screw dislocation along the $z$ axis trapping a giant vortex with $n$ flux quanta in an $s$-wave superconductor. Around the $z$ axis, the phase of the superconducting order parameter winds by $2\pi n$, and we assume a pairing gap of the form $\Delta(\vec{r}) = \Delta_{\sss 0}(r) e^{- i n \varphi}$ in cylindrical coordinates. Schematically, the setup is shown in figure~\ref{fig:schematic}.

In the Nambu basis $\vec{\Psi} = (\Psi_{\uparrow}, \Psi_{\downarrow}, \Psi^{\dag}_{\downarrow}, -\Psi^{\dag}_{\uparrow})^{T}$, the electronic system is described by the Bogoliubov-de~Gennes (BdG) Hamiltonian
\begin{align}
   \mathcal{H}_{\mathrm{BdG}}(\vec{r}) &=\m{H_{0}(\vec{r}) \!+\! H_{\mathrm{soc}}(\vec{r}) &  \Delta(\vec{r}) \sigma_{0} \\ \Delta^{*}(\vec{r}) \sigma_{0} & -\sigma_{y} [H_{0}^{*}(\vec{r}) \!+\! H_{\mathrm{soc}}^{*}(\vec{r})] \sigma_{y}}	\label{eq:BdG}
\end{align}
with the electronic contribution
\begin{align}\label{}
   H_{0}(\vec{r}) &= (\vec{p} + e \vec{A}/c)^{2} / 2m - \EF - g \muB \vec{B}\cdot\boldsymbol{\sigma} / 2,
\end{align}
and spin-orbit contribution \cite{Hu2018}
\begin{align}\label{eq:ScrewSOC}
   \!\!\!
   H_{\mathrm{soc}}(\vec{r}) &\!=\! \alpha(r) (p_{z} / \hbar) [\cos(p_{z} a / 2\hbar) \sigma_{y} \!-\! i\sin(p_{z} a / 2\hbar) \sigma_{x}]\!\!\!
\end{align}
imposed by the screw winding.
Here, we have the momentum operator $\vec{p}=-i\hbar\nabla$, the electron charge $-e \!<\! 0$, the period length $a$ of the screw dislocation, the Fermi energy $\EF$, the gyromagnetic ratio $g$, and the Bohr magneton $\muB$. The magnetic field $\vec{B}$ relates to the vector potential $\vec{A}$ via $\vec{B} = \nabla\!\times\!\vec{A}$. The Pauli matrices $\vec{\sigma} \equiv (\sigma_{x},\sigma_{y},\sigma_{z})^{T}$ act in spin space, while those acting in the particle-hole space are denoted with $\tau_{x,y,z}$. In compounds with high ionicity, the role of $\sigma_x$ and $\sigma_y$ in Eq.~\eqref{eq:ScrewSOC} may be exchanged as shown in Ref.~\cite{Hu2018}. This modification has no implications on our following arguments.

\begin{figure*}[t]
\includegraphics[width = 0.93\textwidth]{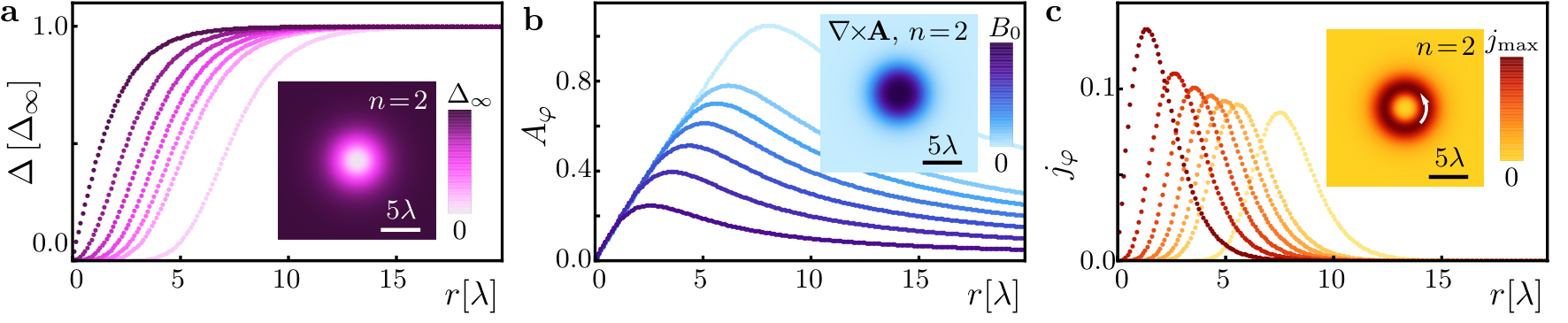}
\caption{
Radial profile of superconducting vortices with $n \!=\! 1\text{-}6,10$ flux quanta (dark to bright).
\textbf{a} superconducting order parameter, inset: planar profile for $n=2$;
\textbf{b} vector potential, inset: planar profile of the magnetic field for $n=2$;
\textbf{c} supercurrent, inset: planar profile for $n=2$.
The dimensionless quantities are defined in \cite{Suppl}.
\label{fig:vortex}
}
\end{figure*}

The Hamiltonian possesses the particular rotational symmetry $[ \mathcal{H}_{\mathrm{BdG}},J_z ] =0$, with the modified angular momentum operator $J_{z} \equiv -i\hbar \partial_{\varphi} + (n/2)\tau_{z}$, which accounts for the vorticity $n$. Note that the translation along $z$ by the distance $a$ (as part of the screw operation) has been omitted in $J_z$ as it does not affect the topological regime, see Supplementary Material \cite{Suppl}.
The eigenvalue $\ell$ of $J_z$ is a good quantum number and the eigenfunctions of $\mathcal{H}_{\mathrm{BdG}}$ may be labeled as
\begin{align}\label{}
   \vec{\Psi}_{\ell}(r,\varphi,z) = \vec{\psi}_{\ell}(r,z) e^{i [\ell - (n/2)\tau_z] \varphi}.
\end{align}
The constraint $\ell - (n/2) \!\in\! \mathbb{Z}$ assures that the wave functions remain single-valued under a circulation around the giant vortex, $\varphi \!\rightarrow \! \varphi \!+\! 2\pi$. This yields integer (half-integer) values for $\ell$ when $n$ is even (odd).

As a minimal requirement, MBS must be eigenstates of the particle-hole operator $\mathcal{C}=\tau_y\sigma_y\mathcal{K}$, with $\mathcal{K}$ the complex conjugation. Since the particle-hole operator anti-commutes with the modified angular momentum, $\{\mathcal{C},J_{z}\}=0$, it holds $\ell \mapsto -\ell$ under $\mathcal{C}$. Consequently, any MBS must have $\ell=0$ to be mapped onto itself. Together with the single-valuedness of $\vec{\Psi}_{\ell}$, it follows that $n\in 2\mathbb{Z}$ is a 
\emph{necessary} condition for the existence of MBS. Similar arguments are known from full-shell nanowires \cite{Vaitiekenas2020} and magnetic skyrmions \cite{YSK16,RGM19}, and go back to earlier work on vortex-bound zero modes \cite{Vol99}.
The system of bound-states centered around the vortex core can be assigned the same topological $\mathbb{Z}_{2}$ index as multi-mode quantum wires \cite{PoL10, LSD11, TeS12, Sam20}, see Supplementary Material \cite{Suppl}. However, topological transitions can only be triggered by a band inversion of modes with $\ell \!=\! 0$.
To gain more qualitative insight, the problem is reduced to a one-dimensional effective problem in the following.

\section{3. Reduction to one dimension}
The angular dependence can be removed from the problem by virtue of the unitary transformation
\begin{align}\label{eq:PhiTrafo}
   U_{\sss J} \equiv e^{-i [\ell - (n/2)\tau_z] \varphi}.
\end{align}
This motivates us to narrow the following discussion to the subspace of even vorticity $n$ and $\ell = 0$. Neglecting the radial dependence of $\alpha(r)$ on the scale of the giant vortex, the wave function decomposes into $\vec{\psi}_{0}(r,z) = \vec{\psi}_{0}(z) \chi(r)$, with a scalar radial part $\chi(r)$ and an axial part $\vec{\psi}_{0}(z)$.
While the separation of variables helps to streamline our analysis, the assumption on $\alpha$ must be relaxed in most real materials, where the Thomas-Fermi length is shorter than the superconducting coherence length. At the same time, our qualitative findings are robust toward a decaying shape of $\alpha(r)$ as long as the net spin-orbit effect on the relevant CdGM wavefunctions does not vanish. For the CdGM states, which have energies well inside the bulk gap $\Delta_\infty$, $\chi(r)$ is localized at the vortex. %
Here, we neglect all other states to study the low-energy properties of the system.
To capture the emergence of MBS, it is convenient to
average out the radial dependence and obtain an effective one-dimensional problem along the $z$ axis. Let the probability-weighted planar mean for a quantity $X(r)$ be
\begin{align}\label{eq:averaging-procedure}
  \inang{X(r)}_\chi
      &= \int_0^\infty dr \, r\, \chi^{\dagger}(r) X(r) \chi(r).
\end{align}
The mean radius $R = \inang{r}_\chi$, superconducting gap $\bar{\Delta}=\inang{\Delta_{\sss 0}(r)}_\chi$, and field strength $\bar{B}=\inang{B(r)}_\chi$ are of particular interest here. Furthermore, the mean Zeeman splitting $\bar{E}_{\sss Z} = g\mu_B\bar{B}/2$ is introduced for brevity. 
Note that even in the presence of in-plane anisotropy, a reduction to a low-energy 1D model of vortex bound states is possible, if an appropriate form factor $f(\varphi)$ is introduced in the planar average. 
After a spin-space rotation with $U_{\sigma} \!=\! \exp[i\sigma_z k_z a/4]$, the effective low-energy 1D Hamiltonian reads
\begin{align}\label{eq:1D-Hamiltonian}
   H_\text{eff} &= \Big[\frac{\hbar^2k_z^2}{2m}-\bar{\mu}\Big]\tau_z
-\bar{E}_{\sss Z}\sigma_z
+\alpha k_z\tau_z\sigma_y
+\bar{\Delta}\tau_x,
\end{align}
with an effective chemical potential $\bar{\mu}\ll\EF$ \cite{Suppl}.
By virtue of equation~\eqref{eq:1D-Hamiltonian}, the electronic states within the vortex/dislocation tube are described by the well-known Hamiltonian of Majorana nanowires, which enters the topological phase for \cite{ORO10, LSD10}
\begin{align}
\label{eq:topol-cond}
\bar{E}_{\sss Z}^{\,2} > |\bar{\Delta}|^2 + \bar{\mu}^2,
\end{align}
i.e., for sufficiently large effective field strength $\bar{B}$.
In this phase, MBS appear at the ends of the wire, that is at the two surface terminations of the vortex/dislocation axis. Remarkably, this demonstrates that the system at hand---a trivial superconductor---is generically susceptible to a topological regime with MBS. To solidify this conceptional result, we give more details on the nature of the transition and suitable material properties below.

\section{4. Giant vortices and CdGM states}
The spatial distribution of the superconducting order parameter $\Delta$ and the magnetic vector potential $\vec{A}$ is well captured within the Ginzburg-Landau (GL) theory.
In reduced units, see references \cite{Blatter1994, Palonen2013} and the Supplementary Material \cite{Suppl}, the free energy density
\begin{align}\label{eq:tildeFGL}
   \!\!\!
   \mathcal{F}_{\sss \mathrm{GL}} = 
      \frac{\kappa^{2}}{4} (|\Delta|^{2} - 1)^{2} +
      \frac{1}{2} \big|(\vec{\nabla} \!+\! i \vec{A})\Delta\big|^{2} +
      \frac{1}{2} (\vec{\nabla} \!\times\! \vec{A})^{2}\!
\end{align}
is solely characterized by the dimensionless quantity $\kappa = \lambda / \xi$, the ratio of the superconducting penetration depth $\lambda$ and the coherence length $\xi$.
Taking full account of the dislocation, the screw symmetry imposes a staircase winding of the supercurrent. However, as shown by Ivlev an Thompson \cite{Ivlev1991}, this effect is parametrically small in $a/\lambda$, and shall be neglected here.
The parameter $\kappa$ is decisive of whether a superconductor is of type-I ($\kappa < \kappa_{c} \equiv 1/\sqrt{2}$) or type-II ($\kappa > \kappa_{c}$). Close to $\kappa_{c}$, the vortex-vortex interactions become small, allowing for stable giant vortices with intermediate vorticities.

Numeric vortex solutions in the weak type-I case ($\kappa = 0.4$), figure~\ref{fig:vortex}, define the confining normal-core $\Delta(\vec{r})$, vector potential $\vec{A}(\vec{r})$, and current $\vec{j}(\vec{r})$ relevant for electronic CdGM states.
Let us stress that type-I superconductors of finite thickness are ready to host vortices with a low number of flux quanta \cite{Huebener2001}. In fact, \emph{single-flux} vortices become stable below a threshold thickness $\sim\!\xi / (1 \!-\!\kappa/\kappa_{c})$.
Close to $\kappa_{c}$, this mesoscopic length bridges from the microscopic scales
of the superconductor to the macroscopic lengths in bulk specimen.
Furthermore, the screw dislocation provides an additional pinning potential, see \cite{Suppl}. Note that a weak type-II superconductor with a strong pinning potential at the dislocation line would be an equally suitable candidate for our proposal.
This gives us good reason to believe that the existence of giant vortices trapped at dislocation lines should persist to specimens of macroscopic size.

\begin{figure*}[t]
\includegraphics[width = 0.93\textwidth]{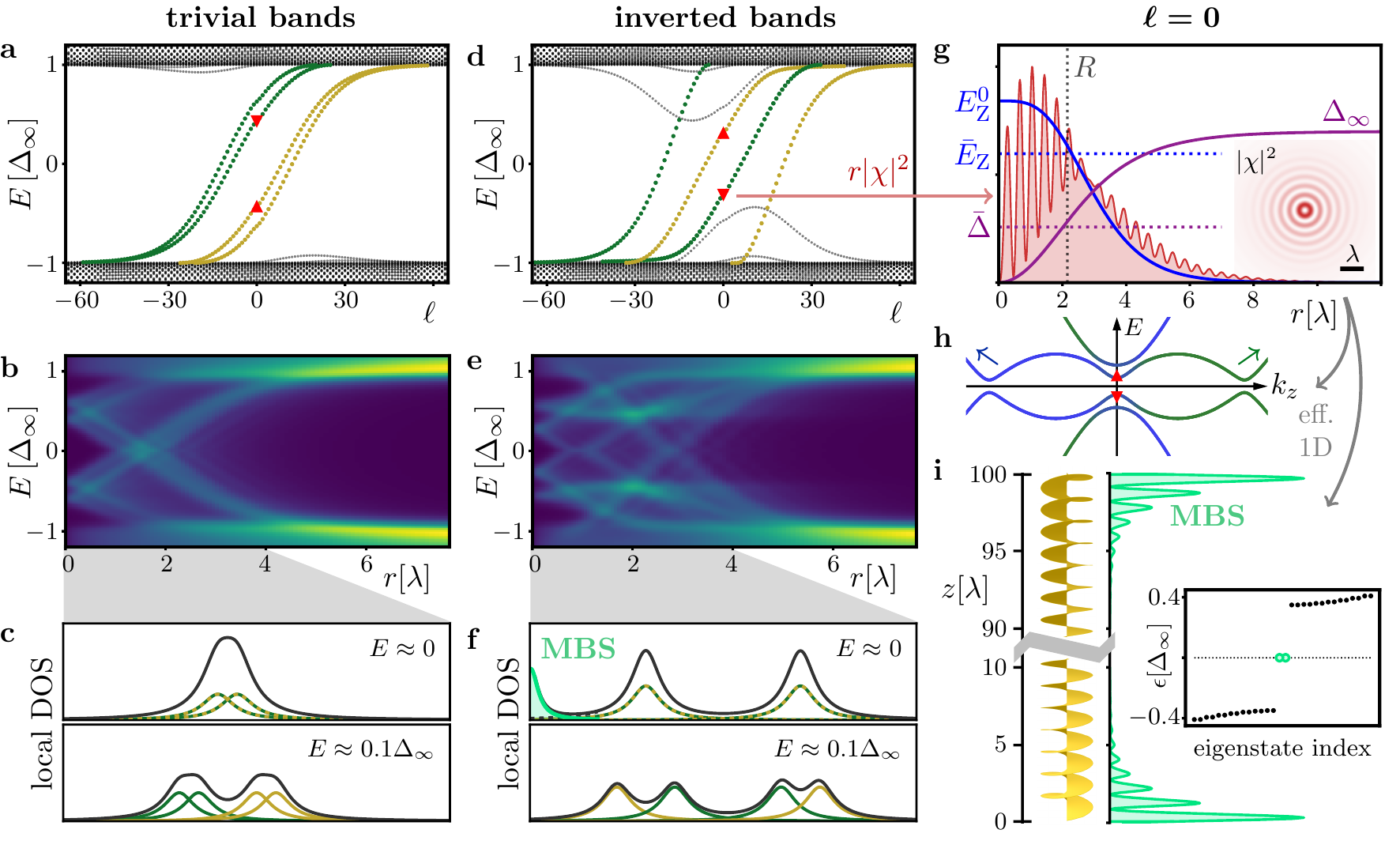}
\caption{
Electronic bound states in a giant vortex with $n \!=\! 2$.
For weak Zeeman splitting, the CdGM band structure remains topologically trivial:
Exact diagonalization reveals the spectrum [\textbf{a}] and the radially resolved density of states (DOS) [\textbf{b}], where the schematic peak structure [\textbf{c}] has no zero-bias features at the core.
When the innermost CdGM bands get inverted by sufficiently strong Zeeman splitting and $n$ is even, a topological regime opens up, cf.\ the spectrum [\textbf{d}]. The DOS [\textbf{e}] reveals a gap at $r=0$, similar to the trivial case [\textbf{a},\textbf{b}]. However, owed to the non-trivial topology, an additional isolated zero-bias peak caused by MBS is expected at the surface termination [\textbf{f}].
The MBS are obtained from an effective 1D wire model, see equation \eqref{eq:1D-Hamiltonian}, for which the radial probability weight $r|\chi(r)|^2$ of the lowest-energy CdGM state at $\ell=0$ (red triangles in \textbf{d}), yields effective (averaged) 1D parameters: $R$, $\bar{\Delta}$, and $\bar{E}_{\sss Z}$ [note also $E_{\sss Z}^0=g\mu_B B(0)/2$]. The planar probability density $|\chi(x,y)|^2$ is shown as an inset.
As a result, the band dispersion [\textbf{h}] in the $z$ direction of the effective 1D system (along the vortex/dislocation axis, color scale represents spin polarization parallel to the spin-orbit field) develops a topological character
and induces zero-energy MBS [\textbf{i}] at the surface terminations (here from exact diagonalization). The inset shows the energies of the lowest eigenstates.
The definitions of dimensionless quantities and all numerical parameters are introduced in the Supplementary Material \cite{Suppl}.
\label{fig:boundstates}
}
\end{figure*}

The confinement of electrons by the superconducting flux tube provides discrete CdGM in-gap states that crucially depend on the vorticity \cite{Caroli1964, Tanaka1993, Volovik1993, Tanaka2002, Virtanen1999, Berthod2005}. These states are obtained by solving the BdG Hamiltonian~\eqref{eq:BdG} in the plane orthogonal to the vortex axis [recall $\alpha(r)$ is assumed constant within the vortex region]. Figure~\ref{fig:boundstates}\textbf{a}-\textbf{f} shows the spectrum of CdGM states and the corresponding density of states.
In absence of Zeeman splitting, the CdGM states disperse as a function of the (discrete) quantum number $\ell$ and a total of $|n|$ spin-degenerate branches cross zero energy \cite{Volovik1993}. As derived in reference~\cite{Berthod2005}, these states have a typical spacing $\Delta E \approx (\Delta_{\sinfty}^{2} / \EF ) \frac{\pi^{2} |n|^{1-\gamma}}{2\pi + 4|n|^{\gamma}}$, with $\Delta_{\sinfty}$ the asymptotic superconducting energy gap and $\gamma$ a non-universal exponent close to unity (reference~\cite{Berthod2005} reports $\gamma \approx 0.78$). Flux tubes with an odd vorticity feature a large number of low-energy states at small $|\ell|$. Giant vortices with an even number of flux quanta, on the contrary, have a large energy gap of the order of $\Delta_{\sinfty}$ at $\ell = 0$. Combining these two results one arrives at a crude, yet surprisingly accurate, estimate for the separation $\Delta\ell \approx 2(\EF / \Delta_{\sinfty}) /|n|^{1-\gamma}$ between branch crossings as a function of the modified angular momentum
$\ell$.

\section{5. Topology and Majorana bound states}
Including the Zeeman effect, the spin degeneracy is lifted and each branch splits into a pair of non-degenerate branches (in the two-fold redundant BdG formalism), with clear signatures in the local density of states. 
Upon increasing the splitting, the two branches closest to zero energy at $\ell=0$ get inverted, which corresponds to the topological transition captured by equation~\eqref{eq:topol-cond}: The CdGM spectrum at $\ell=0$ maps approximately to the spectrum of the one-dimensional Hamiltonian with averaged fields, equation~\eqref{eq:1D-Hamiltonian}, at $k_z=0$.
In figure~\ref{fig:boundstates}, both the trivial CdGM [\textbf{a}-\textbf{c}] and inverted (topological) bands [\textbf{d}-\textbf{f}] are shown.
A mapping to an effective 1D wire is achieved by radial averaging, see figure~\ref{fig:boundstates}\textbf{g} and \cite{Suppl} for details.
The inversion of the lowest two states indicates the gap closing and reopening known from Majorana wires, see figure~\ref{fig:boundstates}\textbf{h} for the dispersion relation for the inverted-band case.
Note that only the central CdGM bands of the 1D system matter for the topological phase and could, in principle, be directly mapped to the Kitaev chain \cite{Kit01} without the detour to the nanowire model. However, to make the role of the spin-orbit coupling more apparent we keep a four-band effective model.
The radial averaging \eqref{eq:averaging-procedure} bears several advantages. On the one hand, it formally maps a higher-dimensional problem to a 1D effective model, see equation \eqref{eq:1D-Hamiltonian}, which is readily identified with the topological nanowire. In the Supplementary Material we provide a detailed argument that the topological character is independent of this reduction scheme. On the other hand, the numerical solution of the problem is now split in two tasks: First, the vortex core and CdGM states are computed for the radial problem, figure \ref{fig:boundstates}\textbf{a}-\textbf{g}. In the second stage, the effective parameters allow to solve the nano-wire problem, figure \ref{fig:boundstates}\textbf{h},\textbf{i}. Both tasks can be performed with high numerical accuracy. The full two-dimensional exact diagonalization calculations in the $rz$-plane---a task that becomes necessary if the spin-orbit coupling shows a strong spatial dependence---would impose severe numerical limitations on the system size, and the results would be overly prone to finite-size errors.

In the topological phase, MBS appear at the terminations of the wire, i.e.\ one at each sample surface. The CdGM states can be resolved in the differential conductance of scanning tunneling experiments \cite{YPH21}, as shown schematically in figure~\ref{fig:boundstates}\textbf{c},\textbf{f}. A MBS would lead to an \emph{additional} peak located at $E=0$ and close to $r=0$. Such a peak---unattached to the dispersive CdGM bands---is a measurable signature of the MBS. The relative strength of this peak in comparison to the other subgap states depends on the localization of the MBS at the surface. The probability density of the MBS along the $z$ axis is shown in Figure~\ref{fig:boundstates}\textbf{i}. The characteristic length $\zeta\approx v_F/\Delta_\text{eff}$ of the exponential localization $\propto e^{-z/\zeta}$ is inversely proportional to the effective gap $\Delta_\text{eff}$ in 1D. The latter may be given by the band gap at either $k_z=0$ or $k_z\approx \kF$, with the Fermi wave vector $\kF$ of the effective wire.

\section{6. Material requirements}
Before concluding, let us briefly comment on the energy scales involved in the problem, and how they cooperate to favor a topological phase:
The five independent energy scales to consider here are
(i) the kinetic energy $K = \hbar^{2} / 2m \xi^{2}$ at wavelengths $\sim\!\xi$, depending on the \emph{effective} mass $m$,
(ii) the Fermi energy $\EF$
(iii) the bulk superconducting gap $\Delta_{\sinfty}$,
(iv) the Zeeman splitting $\bar{E}_{\sss Z}$,
and (v) the spin-orbit splitting $\alpha \kF$ induced by the screw dislocation.
Apart from natural constants, the Zeeman splitting only depends on the superconducting penetration depth, $\bar{E}_{\sss Z}\propto\lambda^{-2}$ (assuming $\kappa\approx1$), and is estimated to be in the range $10\text{-}100\,\mu\mathrm{eV}$ for typical values of $\lambda$. To reach the topological regime, the spacing of the lowest CdGM states at $\ell \!=\! 0$ must not exceed the Zeeman energy. At the same time, the superconducting gap $\Delta_{\sinfty}$ needs to be large enough to allow for measurements. As the spacing between CdGM branches decreases with $K$, a large effective mass would facilitate the band inversion. Furthermore, $\EF/\Delta_{\sinfty}$ should be moderate to ensure an unambiguous resolution of the MBS zero-energy peak. Also, our effective low-energy model is only valid for a small Fermi energy.
This is again favored by a large effective mass. Spin-orbit effects enter the game once band inversion is present: opening a substantial gap at $k_z = \kF$ reduces the localization length $\zeta$ and avoids the hybridization of the two Majorana bound states.
Spin-orbit coupling is guaranteed by the dislocation~\cite{Hu2018}; the strength depends on the chemical composition of the material and will likely increase for heavier elements. Whereas our approximation of a radially constant parameter $\alpha(r)\approx\alpha$ requires the Thomas-Fermi length to be comparable to the coherence length, relaxing the approximation will not qualitatively affect the findings, but merely add quantitative constraints.
In summary, we expect our proposal to apply to superconductors with $\kappa\approx1$, large effective mass $m$, not too strong superconducting gap $\Delta_\infty$, and a crystal system known to host screw dislocations. Suitable giant vortices can be created by sweeps of an external magnetic field.

Identifying a specific material that fulfills all of the requirements above is a challenging task that goes beyond the scope of this work. Although we cannot predict a strong candidate, we conclude our discussion with a brief prospect on relevant material classes for future investigations related to our conceptual proposal. We focus on two key properties: the capability to host multi-quantum vortices and a small Fermi energy. The former by itself is not too difficult. With a rapidly increasing number of reported compounds \cite{Sun2016, Singh2018, Peets2019, GarciaCampos2021}, finding superconductors at the type-I/type-II-crossover is very realistic. Alternatively, giant vortices can be created in type-II superconductors by means of lateral specimen confinement and control of vortex pinning \cite{Grigorieva2007, Cren2011, Ge2016, Golod2019}.
Exceptionally small Femi levels, on the other hand, have recently been reported in iron-based compounds such as FeSe \cite{Coldea2018, Huang2021}. In addition, there is a growing family of superconducting-doped semiconductors \cite{Bus15} with members of both type-I and type-II. Such materials frequently have unconventional properties that are absent in our model, and doping would also require to account for disorder effects. More specialized models combined with DFT results may help to find a good candidate in future work.

\section{7. Conclusion}
We demonstrated that a pair of topological lines---a giant vortex trapped at a screw dislocation---is capable of creating a quasi-one-dimensional topological wire hosting MBS in an otherwise trivial superconductor. This proposal now awaits physical realization, which---unlike other defect-based systems, e.g. \cite{CJZ20}---is facilitated by the trivial bulk topology of the superconducting state. Screw dislocations are ubiquitous and their appearance can often be controlled during sample growth. Giant vortices, on the other hand, are known to exist in easily available materials. Therefore, our proposal opens an exciting and realistic route to hunt for MBS without sophisticated fabrication of nano-hybrid structures, such that topological superconductivity may become accessible for a broad experimental community.

\section*{Acknowledgments}
We are extremely grateful to T.\ Gozlinski, Q.\ Li, and W.\ Wulfhekel, who provided the initial spark to this project and challenged it with critical interactions and discussions. We thank V.B.\ Geshkenbein, I.V.\ Gornyi, 
F.\ Hassler, R.P.\ Huebener, A.D.\ Mirlin, and J.\ Schmalian for useful discussions.
S.R. was supported by the Deutsche Forschungsgemeinschaft via the Grants No. MI 658/12-1 (joint DFG-RFBR project) and No. MI 658/13-1 (joint DFG-RSF project). %
R.W. acknowledges funding support from the Heidelberger Akademie der Wissenschaften through its  WIN initiative (8. Teilprogramm).

\noindent

\begin{widetext}
\end{widetext}
\appendix
\section{SUPPLEMENTARY MATERIAL}

\emph{Free energy of vortices}---The radial dependence of the gap function $\Delta_{\sss 0}(r)$ and the vector potential $A_{\varphi}(r)$ can be evaluated numerically by minimizing the Ginzburg-Landau (GL) free energy density. Starting from the general expression \cite{Blatter1994}
\begin{align}
   \mathcal{F}_{\sss \mathrm{GL}} &= 
      \alpha |\Delta(\vec{r})|^{2} + \frac{\beta}{2} |\Delta(\vec{r})|^{4}
			+ \frac{1}{8\pi} [\vec{\nabla} \times \vec{A}(\vec{r})]^{2}
      \nonumber\\ 
      &\quad
      +\!
      \frac{\hbar^{2}}{2 m} \big|[\vec{\nabla} + i (2e/\hbar c) \vec{A}(\vec{r})]\Delta(\vec{r})\big|^{2}
			\label{eq:FGL}
\end{align}
and introducing the dimensionless quantities
$\tilde{\Delta} \!=\! \Delta / \Delta_{\sinfty}$,
$\kappa^{2} \!=\! (m^{2}c^{2}\beta / 8 \pi e^{2} \hbar^{2})$, 
$\mathcal{\tilde{F}}_{\sss \mathrm{GL}} \!=\! (\kappa^{2}/4)[1 \!+\! (2\beta / \alpha^{2}) \mathcal{F}_{\sss \mathrm{GL}}]$,
$\vec{\tilde{r}} \!=\! (16 \pi e^{2} |\alpha|/m c^{2} \beta)^{1/2} \vec{r}$, and
$\vec{\tilde{A}} \!=\! (m \beta / 4 \pi \hbar^{2} |\alpha|)^{1/2} \vec{A}$, with $\Delta_{\sinfty} \equiv (\beta / |\alpha|)^{1/2}$ we arrive at equation \eqref{eq:tildeFGL}. Note that in the main text tildes are dropped for simplicity.  The supercurrent density $\vec{j} \propto |\Delta|^{2} \vec{A}$ and the field distribution $\vec{B} = \nabla \!\times\! \vec{A}$ are directly derived from the primary quantities. The expression for $\vec{j}$ considers the present cylindrical geometry.
Using further the rotational symmetry of the problem and that $\Delta(\vec{r}) = \Delta_{\sss 0}(r) e^{-i n \varphi}$, the solution to the GL equation reduces to minimizing
\begin{align}\label{eq:fullGL}
   \!\!
   \mathcal{E}_{\sss \mathrm{GL, 1D}} &= \int_{0}^{\infty} \tilde{r} d\tilde{r} \
      \frac{\kappa^{2}}{4} (\tilde{\Delta}_{\sss 0}^{2} \!-\! 1)^{2}
      + \frac{1}{2} (\partial_{\tilde{r}}\tilde{\Delta}_{\sss 0} )^{2}
      \nonumber\\
      &\qquad\quad
      + \frac{1}{2 \tilde{r}^{2}} \big\{
         \big[ (\tilde{r} \tilde{A}_{\varphi} \!-\! n )\tilde{\Delta}_{\sss 0}\big]^{2} + \big[ \partial_{\tilde{r}} (\tilde{r} \tilde{A}_{\varphi}) \big]^{2} \big\}\!
\end{align}
for specified $\kappa$ and winding $n$. Here $\tilde{\Delta}_{\sss 0} = \Delta_{\sss 0} e^{i n \varphi}$ is a real scalar function of the radial coordinate only. Our numerical minimization follows a stochastic annealing scheme.
As the planar profile of the vortex at a screw dislocation remains nearly unaffected by the screw symmetry \cite{Ivlev1991}, this effect is neglected here. On the other hand, the interaction between the giant vortex and the defect may define a relevant pinning energy. If desired, this effect can be accounted for, e.g.\ by a spatial modulation of $T_{c}$: With an energy distribution $e_{\mathrm{pin}}(\tilde{r}) = (\kappa^{2}/2) \delta T_{c}(r) |_{r = \lambda \tilde{r}}$ equation \eqref{eq:fullGL} is augmented by
\begin{align}
   \mathcal{E}_{\mathrm{pin}} = \int_{0}^{\infty} \tilde{r} d\tilde{r} \ e_{\mathrm{pin}}(\tilde{r}) \big[\tilde{\Delta}_{\sss 0}(\tilde{r}) \big]^{2}.
\end{align}

\medskip

\emph{Effective BdG Hamiltonian}---From the functions $\Delta_{\sss 0}(r)$ and $A_{\varphi}(r)$, we obtain the spectrum and radial wave functions of the CdGM states for the Hamiltonian \eqref{eq:BdG}. After the unitary transformation, equation~\eqref{eq:PhiTrafo}, we have
\begin{align}
   \tilde{\mathcal{H}}_{\mathrm{BdG}} &\equiv U \mathcal{H}_{\mathrm{BdG}} U^{\dag} \nonumber\\
      &= \!\big\{\!
         \big(\frac{p_{r}^{2}\!+\!p_{z}^{2}}{2m}\!-\! \EF \big)
         +\frac{\hbar^{2}}{\!2m r^{2}\!}\big[\ell \!-\! \big(\frac{n}{2} \!-\! \frac{e A_{\varphi} r}{\hbar c}\big) \tau_z\big]^{2}
    \big\} \tau_z
	\nonumber \\
    &\quad
    + \alpha(r) \frac{p_{z}}{\hbar} \big[\cos\left(\frac{p_{z} a}{2\hbar}\right) \sigma_y - i\sin\left(\frac{p_{z} a}{2\hbar}\right) \sigma_x \big] \tau_z
    \nonumber \\ 
	&\quad
    -\frac12 g\muB B(r) \sigma_z +\Delta_{\sss 0}(r) \tau_x
		\label{eq:FullHamiltonian}
\end{align}
where $p_{r}^{2} = -\hbar^{2}\frac{1}{r} \partial_{r}[r \partial_{r} ]$, $p_{\varphi} = -i \hbar\frac{1}{r} \partial_{\varphi}$, and $p_{z} = -i\hbar\partial_{z}$. 
With constant $\alpha(r) \approx \alpha$ within the vortex core region, separation of variables becomes possible. Then, the second line in equation \eqref{eq:FullHamiltonian} along with the kinetic $\propto p_z^2$ term do not appear in the radial problem. 
It is convenient to absorb the functional determinant $r$ that appears under the integral, $\int r\,dr$, into the radial wave function, $\phi(r)=\sqrt{r}\chi(r)$. The eigenproblem for $\phi(r)$ is solved by exact diagonalization of the Hamiltonian (using the dimensionless quantities defined above) 
\begin{align}
   \!
   H_\phi(r) &\!=\!
      - \frac{K}{\kappa^2} (\partial^2_{\tilde{r}} + 1/4) \tau_z
      \!-\! \EF \tau_z
      \!-\! G \tilde{B}(\tilde{r}) \sigma_z
      \!+\! \Delta_{\sinfty} \tilde{\Delta}(\tilde{r}) \tau_x
       \nonumber
   \\
      &\quad
      +\! \frac{K}{\kappa^2\tilde{r}^2}\Big\{\!
         [\tilde{A}(\tilde{r})\tilde{r}-n] \ell
         + \big[(\tilde{A}(\tilde{r})\tilde{r}-n)^2/4 + \ell^2\big] \tau_z
         \!\Big\},
\end{align}
with the energy $K \!=\! \hbar^2/(2m \xi^2)$ and the Zeeman factor $G \!=\! g\hbar^2/(8 m_e \lambda^2)$,
where $m_e$ is the bare electron mass that enters through $\muB$.
In the calculations shown in figure~\ref{fig:boundstates}, we used the numerical parameters $K/\kappa^2=0.15\,\Delta_{\sinfty}$, $\EF=10\,\Delta_{\sinfty}$, $(\Phi_0/2\pi\lambda^2)G=0.45\,\Delta_{\sinfty}$ (trivial), $(\Phi_0/2\pi\lambda^2)G=3.6\,\Delta_{\sinfty}$ (topological), $\alpha/\lambda=\Delta_{\sinfty}$, and $\kB T=\Delta_\infty/25$ (only for thermal broadening in panels \textbf{b},\textbf{e}).

One may wonder whether the procedure above is accurate when respecting the full screw symmetry rather than cylindrical symmetry. In fact, our Hamiltonian Eq.~\eqref{eq:BdG} accounts for the dislocation only in $H_{\mathrm{soc}}$ and neglects terms that explicitly break the rotational symmetry $J_z$. As mentioned in the main text, full account of the screw symmetry is given when $J_z$ is replaced by $J_zT_z$ (eigenvalue $\ell^\prime$), with a translation operator $T_z$. The Hamiltonian~\eqref{eq:FullHamiltonian} then acquires planar-axial mixing terms involving products of $\ell^\prime$ and $p_z$ and does not allow to solve the radial and axial problems independently. Nevertheless, MBS would still require $\ell^\prime=0$, where the mixing terms vanish. Hence, all corrections invoked by $T_z$ are irrelevant for the topological regime. At larger $\ell^\prime$, the corrections to the CdGM states are negligible at $r\gg a$.

The effective chemical potential in equation~\eqref{eq:1D-Hamiltonian} is
\begin{align}
   \bar{\mu} &= \EF - \frac{\hbar^2}{2m}\Big[
      \inang{(p_r / \hbar)^2}_\chi
      + \frac{n^2}{4} \inang{r^{-2}}_\chi
      \nonumber\\ 
      &\qquad\qquad
      + \Big(\frac{\pi}{2\Phi_0}\Big)^2\inang{r^2B(r)^2}_\chi - \frac{n\pi}{2\Phi_0}\bar{B}\Big].
\end{align}
The dispersion and the MBS shown in figure~\ref{fig:boundstates}\textbf{h},\textbf{i} have been calculated with parameters drawn from the state shown in panel \textbf{g} using our planar averaging procedure, see equation \eqref{eq:averaging-procedure}. With our choice of parameters, we find that $\bar{\mu}\ll \EF$.
This condition reflects the reduction to an effective one-dimensional chain and brings the topological regime within reach, see equation \eqref{eq:topol-cond}. Its realization is facilitated by a low carrier density. Also, we suppose that a strong crystal anisotropy can substantially distort the semiclassical orbits of the CdGM bound states, thereby significantly reducing the cross-section of the effective axial system and its effective chemical potential.

The dispersion was obtained analytically from equation~\eqref{eq:1D-Hamiltonian}, whereas the states in panel \textbf{i} have been calculated by exact diagonalization of the tight-binding version of the same equation on a finite grid (1500 sites).

\medskip

\emph{Dimensionality and topological transition.} In the main text, we employed a reduction scheme to derive an effective one-dimensional model and retrieve the parameters relevant to the appearance of topologically protected Majorana modes. Here, we argue that the topological character of the system is independent of this reduction. Without any simplifications, we study a three-dimensional system with an in-plane confinement exerted by the vortex potential, cf. Fig.~\ref{fig:vortex}. This potential well generates a series of bound states (the CdGM states) with energies inside the bulk gap of the superconductor. As these states are unrestricted in the $z$ direction, the in-gap system is exactly represented as a quantum wire with multiple modes, labeled by the quantum number $\ell$. We note that in anisotropic systems, $\ell$ remains an integer quantum number while deviating from the usual interpretation of an angular momentum quantum number. The $\mathbb{Z}_2$ invariant $Q$ indicating the topological phase of multi-mode superconductor wires follows from the Pfaffian $\mathrm{Pf}$ at the high-symmetry points of the Brillouin zone, $k_z=0$ and $k_z=\pi/a$ \cite{Kit01, PoL10, LSD11, TeS12, Sam20}. In our basis, the Hamiltonian has the particle-hole symmetry operator $\mathcal{C}=\tau_y\sigma_y\mathcal{K}$, thus
\begin{equation}
Q = \text{sgn}\bigg\{\frac{\mathrm{Pf}[\mathcal{H}_\mathrm{BdG, in-gap}(k_z=\pi/a)\tau_y\sigma_y]}
                         {\mathrm{Pf}[\mathcal{H}_\mathrm{BdG, in-gap}(k_z=0)\tau_y\sigma_y]}\bigg\}.
\end{equation}
This $\mathbb{Z}_2$ invariant measures a band inversion at $k_z=0$ relative to $k_z=\pi/a$. It relates to the winding number $\nu\in\mathbb{Z}$ of the chirally symmetric Kitaev chain model via $Q=(-1)^\nu$. In our system, the only source of change of $Q$ are the modes at $\ell=0$: Any band-inversion transition at $\ell\neq0$ would be mirrored by the states at $-\ell$ and thereby not change the sign of the Pfaffian. This is a consequence of our symmetry argument in the main text, $\{\mathcal{C},J_z\}=0$. On the other hand, it is readily clear that a band inversion at $\ell=0$ will flip the sign of $Q$. This band inversion can be driven by the Zeeman field, as one can see from Fugures~\ref{fig:boundstates}\textbf{a},\textbf{d}. Thus, the topological character of the multi-vortex system does not rely on the dimensional reduction scheme. It can rather be deduced from the inverted CdGM spectrum, i.e.\ using the full 3D Hamiltonian.

\end{document}